\documentstyle[nato,numreferences,epsfig]{crckapb}
\begin{opening}
\title{Cosmological implications of massive neutrinos}
\author{L.A. Popa}
\institute{Institute for Space Sciences\\
           R-76900 Bucharest - M\u{a}gurele, Romania}
\end{opening}
\begin{document}
\section{Introduction}

The atmospheric neutrino results reported by the  Super-Kamiokande \cite{sk}
and
MACRO \cite{mac,mac2k} experiments indicate that neutrinos oscillate,
these data being consistent with $\nu_{\mu} \leftrightarrow \nu_{\tau}$
oscillations. The small value of the difference of the squared masses
($5 \times 10^{-4}{\rm eV}^2  \leq \Delta m^{2}
\leq 6 \times 10^{-3}{\rm eV}^2$)
and the strong mixing angle ($ \sin^{2} 2\theta \geq 0.82$) suggest that
these neutrinos are nearly equal in mass
as predicted by many models of particle physics beyond the
standard model.
Also, the LSND experiment \cite{Ath98} supports
$\nu_{\mu}\leftrightarrow \nu_{e}$ oscillations
($\Delta m^{2} \leq 0.2$eV$^{2}$);
some solar neutrino experiments
\cite{Bah98} suggest that $\nu_e$ could oscillate to a sterile
neutrino $\nu_e \leftrightarrow \nu_s$
($\Delta m^{2} \simeq 10^{-5}$eV$^2$).
The direct implication of neutrino oscillations is the existence of
a non-zero neutrino mass in the eV range or lower, and consequently
a not negligible hot dark matter (HDM) contribution $\Omega_{\nu}~\neq 0$
to  the total energy density of the universe.
Recent works \cite{Primack95,Gaw98,Gaw2000} show
that the addition of a certain fraction
of HDM component to the total energy density
of the universe can lead to the agreement between the
Cosmic Microwave Background (CMB) anisotropy power spectrum
at the small scales and  the observations
of the Large Scale Structure (LSS) of the universe as derived by
the redshift galaxy surveys \cite{Primack95}.

 The CMB anisotropy pattern contains information related to
the physical processes
occurring before the last scattering of the CMB photons; the
LSS data reflects the clustering regime effects in our local universe.
These measurements can provide
independent probes for the structure of the universe on similar
comoving scales at different cosmological epochs.
This paper discusses the implications of massive neutrinos
properties for CMB and LSS measurements.

\section{Properties of the neutrino background}

\subsection{Cosmological limits on non-degenerate neutrino masses}

Evidences has been accumulated that our universe
is a low matter density universe (see e.g \cite{Fuk99}
and the references therein).
Indications like the Hubble diagram of Type 1a
supernovae  and
the acoustic peak distribution in the CMB anisotropy power spectra
(see e.g. \cite{Efs99} and the references therein) point to a
universe dominated by vacuum energy (cosmological constant
$\Lambda$) that keeps the universe close to flat.
The combined analysis of the latest CMB anisotropy data
and Type 1a supernovae data
indicates $\Omega_m=0.25^{+0.18}_{-0.12}$ and
$\Omega_{\Lambda}=0.63^{+0.17}_{-0.23}$ \cite{Efs99}
for matter and vacuum  energy densities respectively,
inferring a Hubble constant value of H$_0\approx$
65~km~s$^{-1}$~Mpc$^{-1}$
($h=H_0/100$ km s$^{-1}$Mpc$^{-1}=0.65$).
Adding a HDM component to the $\Lambda$CDM model
($\Lambda$CHDM) leads to a worse fit to LSS data,
resulting in a limit on the total non-degenerated neutrino mass
of $m_{\nu} \leq 2$eV
for a primordial scale invariant power spectrum
and $m_{\nu} \leq 4$eV for a primordial scale free power spectrum
\cite{Gaw2000}.
A stronger upper limit
is obtained  from
the matching condition of the LSS
power spectrum normalization $\sigma_8$ (defined as the {\it rms} amplitude
of the galaxy power spectrum in a sphere of radius 8$h^{-1}$Mpc)
at the COBE scale and at the cluster scale  \cite{Fuk99}.
For the case of the $\Lambda$CHDM model having
$\Omega_m=0.3$, $\Omega_{\Lambda}=0.7$ and a
primordial scale invariant power spectrum it is found an upper limit
of $m_{\nu} < 0.6$  if $H_0 < 80$ km s$^{-1}$ Mpc$^{-1}$.
Also, the constraints on the
cosmological parameters
obtained by using the most recent CMB
anisotropy data \cite{Lange2000}
combined with Type 1a supernovae data implies a best
fit model with
$\Omega_m \approx$ 0.33, $\Omega_{\Lambda}$=0.67 and
a neutrino density parameter
$\Omega_{\nu} \approx$ 0.12 ($\sum_i m_{\nu_i}$
$\approx 92h^2\Omega_{\nu}=4.66$eV)
if priors H$_0$=65 km s$^{-1}$ Mpc$^{-1}$ and $\Omega_{b}$h$^2$=0.02
are assumed \cite{Teg2000}.

\subsection{Cosmological limits on  neutrino degeneracy}

A way to improve the agreement of the cosmological
models involving neutrinos
with the higher values of Hubble parameter is
to consider that neutrinos are degenerated.
The neutrino degeneracy parameter $\xi_{\nu}$
is defined as $\xi_{\nu}=\mu_{\nu}/T_{\nu}$,
where $\mu_{\nu}$ is the neutrino chemical
potential and $T_{\nu}$ is the neutrino temperature.
The neutrino degeneracy enhances the contribution
of the HDM component
to the total energy density of the universe \cite{Lar95}
changing the neutrino decoupling temperature
\cite{Fre83}, the abundance of light elements
at the big bang nucleosynthesis (BBN)
\cite{Kang92}, the CMB anisotropies and the
matter power spectrum \cite{Kin99}.
Constraints on neutrino degeneracy coming from BBN \cite{Kang92}
indicate $-0.06 \leq \xi_{\nu_e} \leq 1.1$
and  $|\xi_{\nu_{\mu,\tau} }| \leq 6.9$.
Bounds on the neutrino degeneracy parameter are obtained
by combining the current observations of the CMB anisotropies and
LSS data \cite{Kin99}.
The analysis of the CMB anisotropy
data obtained by the Boomerang experiment
indicates bounds on massless neutrino degeneracy parameter
of
$|\xi_{\nu_{e,\mu,\tau}}| \leq 3.7$, if only one massless
neutrino species is degenerated, and  $|\xi_{\nu}| \leq 2.4$, if the
asymmetry is equally shared among three
massless species \cite{Han2000}.\\

\subsection{Massive neutrino mixing and the phase space
distribution function}

In the standard assumption of the mixing of massive neutrinos,
the neutrino flavor eigenstates are described by a
superposition of the mass eigenstate components
that propagate differently with different
energies, momenta and masses \cite{Pdg98}.
This paper considers two massive neutrino  flavors
 $\nu_{\mu}$ and $\nu_{\tau}$ and a third neutrino
flavor $\nu_{e}$ is massless.
The mixing occurring in vacuum between
$\nu_{\mu}$ and $\nu_{\tau}$
can be written as:
\begin{eqnarray}
| \nu_{\mu} \rangle & = &\;\;\cos \theta_0 | \nu_{2} \rangle + \sin
\theta_0 | \nu_{3} \rangle \\
| \nu_{\tau} \rangle &= &-\sin \theta_0 | \nu_{2} \rangle + \cos \theta_0
| \nu_{3} \rangle,
\nonumber
\end{eqnarray}
where  $\nu_2$ and $\nu_3$ are the mass eigenstate components and
$\theta_0$ is the vacuum mixing angle.
It is usual to consider that each neutrino
of a definite flavor is dominantly one mass eigenstate \cite{Pdg98}.
In this circumstance, the dominant mass eigenstate
component of $\nu_{\mu}$
is $\nu_2$, that of $\nu_{\tau}$
is $\nu_3$
and their difference of the squared masses is
$\Delta m^2=m^2_3-m^2_2$.

In the expanding universe, the system of neutrinos
drops out from thermal equilibrium with $e^{\pm}$,
photons, the small fraction of baryons and other massive species
(e.g. $\mu^+ \mu^-$ pairs) when the ratio of the averaged weak
interaction rate to the expansion rate falls below unity.
The present values of neutrino temperature $T_{\nu_0}$
and the present temperature of the CMB photons
$T_0 \approx 2.725$K
are related through  $T_{\nu_0}/T_{0}\simeq (3.9/g_{*D})^{1/3}$,
where $g_{*D}$ is the number
of degrees of freedom in equilibrium at
neutrino decoupling.
For a degeneracy parameter
$\xi_{\nu} \leq 15$ the neutrino decoupling temperature
is of few MeV  \cite{Fre83} and
from the entropy conservation one obtains that after decoupling
$T_{\nu_0}\simeq (4/11)^{1/3}T_0$.
It follows that the total degeneracy parameter
$\xi_{\nu}$ is conserved and the present lepton asymmetry
of the neutrino background is \cite{Kang92}:
\begin{eqnarray}
L_{\nu}=\frac{1}{12 \zeta(3)} \frac{T_{\nu_0}}{T_0}
[\xi_{\nu}^3+\pi^2 \xi_{\nu}], \nonumber
\end{eqnarray}
where $\zeta(3)$ is the Riemann function of 3.
At decoupling, neutrino
with   mass in eV range
behave like a relativistic particle
and its full phase space distribution function can
be written as a pure Fermi-Dirac distribution function
$f^{0}_{\nu}(q)$,
a zero-order distribution that depends only
on momenta, plus a perturbed part $\Psi({\bf x},{\bf q},a)$
that depends on momenta, position and time \cite{MaB95}:
\begin{eqnarray}
f_{\nu}({\bf x},{\bf q},a)=f^{0}_{\nu}(q)(1+\Psi({\bf x},{\bf q},a))
\hspace{1cm}
f^{0}_{\nu}(q)=\frac{1}{{\rm e}^{E_{\nu} / T_{\nu}\pm \xi_{\nu}}+1}\,,
\end{eqnarray}
where:  $E_{\nu}=\sqrt{q^2+
a^2 m^{2}_{\nu}}$ is the neutrino energy,
$a$ is the cosmological scale factor
($a_0=1$ today), $q$ is the comoving neutrino momentum,
$\pm \xi_{\nu}$ is the neutrino(-sign)/antineutrino(+sign)
degeneracy parameter
and $m_{\nu}$ is the dominant mass component.
The full neutrino phase space distribution function
was computed
through numerical simulations \cite{Popa2001}
based on the standard particle-mesh (PM) method.
The initial neutrino positions and velocities
was generated from the HDM matter density fluctuation power spectrum
by using Zel'dovich approximation \cite{Zel70}.
Figure 1 shows (left panels)
the contour plots of the constant particle probabilities
in  $\delta q$-$\delta \rho_{HDM}$ plane,
where $\delta q$ is the fluctuation of the neutrino comoving momentum
and $\delta \rho_{HDM}$ is the fluctuation of
the neutrino density field.
It is also shown (right panels)
the momentum distribution
functions obtained from numerical simulations
(continuous line)
compared with a pure Fermi-Dirac distribution function
(dashed line).
\begin{figure}
\vspace{-0.6cm}
\epsfig{figure=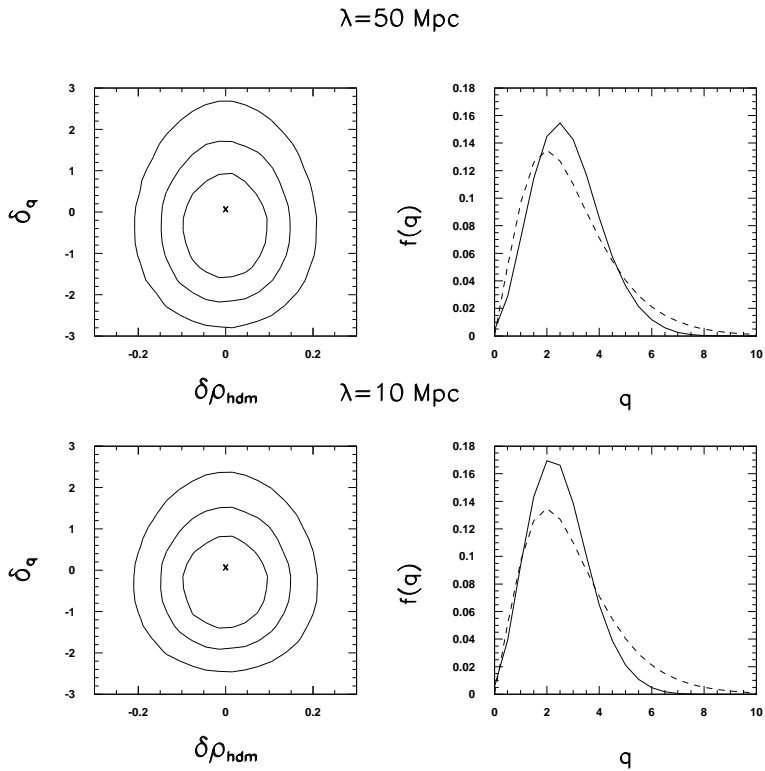,height=11cm}
\caption{Left panels: the contour plots of constant particle probabilities
in the
$\delta q - \delta \rho_{HDM}$ plane. From exterior to
interior the contours correspond
to: 0.75, 0.5 and 0.25 probability.
Right panels: the full neutrino phase space distribution functions
obtained from numerical simulations (continuous line) and
the pure Fermi-Dirac distribution function (dashed line).}

\vspace{-0.5cm}
\end{figure}
The time evolution of the phase space distributions
$f_{\nu_{\mu}}$ and $f_{\nu_{\tau}}$
of $\nu_{\mu}$
and $\nu_{\tau}$  can be written as \cite{Popa2000}:
\begin{eqnarray}
\partial f_{\nu_{\mu}} ({\bf x},{\bf q},t)/\partial t
& = &
-< {\cal P} >_{\nu_{\mu}}  f_{\nu_{\mu}}({\bf x},{\bf q},t)+
< {\cal P } >_{\nu_{\tau}}  f_{\nu_{\tau}}({\bf x},{\bf q},t),
\nonumber \\
\partial f_{\nu_{\tau}}({\bf x},{\bf q},t)/\partial t
& = &
-< {\cal P} >_{\nu_{\tau}}  f_{\nu_{\tau}}({\bf x},{\bf q},t)+
< {\cal P } >_{\nu_{\mu}}  f_{\nu_{\mu}}({\bf x},{\bf q},t),
\end{eqnarray}
where: $< {\cal P } >_{\nu_{\mu}}$ and $< {\cal P } >_{\nu_{\tau}}$
are the neutrino mixing probabilities,
$dt=da/(aH)$ and $H$ is the Hubble expansion rate:
\begin{eqnarray}
H^2=\frac{8 \pi G}{3}[\Omega_m/a^3+\Omega_r/a^4+
\Omega_{\Lambda}+\Omega_k/a^3].
\end{eqnarray}
In the above equation $G$ is the gravitational constant,
$\Omega_m=\Omega_b+\Omega_{cdm}+\Omega_{\nu}$
is the matter energy density parameter and $\Omega_b$, $\Omega_{cdm}$,
$\Omega_{\nu}$ are the energy density parameters for baryons, cold dark matter
and neutrinos, $\Omega_r$ is the energy density parameter for radiation
(including photons and one relativistic neutrino flavor),
$\Omega_{\Lambda}$ is the vacuum energy density parameter and
$\Omega_k=1-\Omega_m-\Omega_{\Lambda}$ is the energy density parameter
related to the curvature of the universe.
The set of equations (3) assume that,
apart from the neutrino mixing, no other physical
processes take place, leading to the conservation of
total neutrino
number densities $n_{\nu}$ and degeneracy parameters
$\xi_{\nu}$. It follows that
non-vanishing  mixing and degeneracy parameters
would lead to changes of the
energy density, pressure, their perturbations, the shear stress
and the energy flux of  each massive neutrino flavor,
quantities that are related to the neutrino phase space distribution
function \cite{Popa2000}.
Figure ~2 presents the time evolution of the energy density parameters
and of the density perturbations of different
components (matter, radiation, vacuum, neutrinos)
for few values of neutrino mixing and degeneracy
parameters. One can observe that the variation
of $\Delta m^2$ and $\Delta \xi_{\nu}$
affect the energy density parameters and
the density perturbations for all components, while
the variation of $\sin^2 2\theta_0$ affect only
the neutrino density perturbations, reflecting the
conservation conditions imposed by equations (3).
\begin{figure}
\vspace{-2cm}
\epsfig{figure=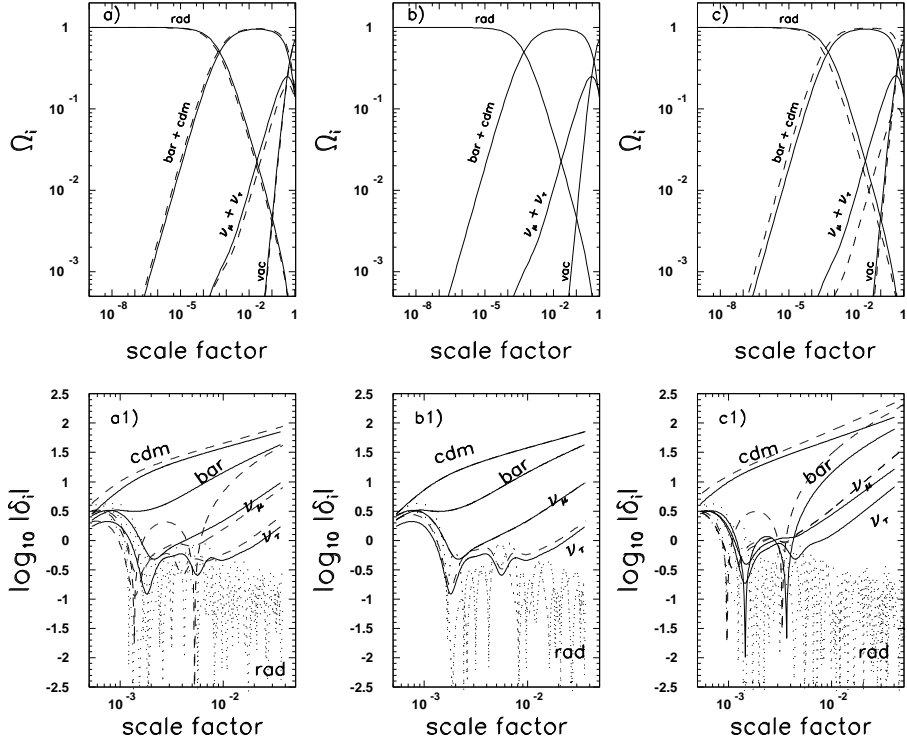,height=13cm}
\caption{The evolution with the scale factor of the energy
density parameters $\Omega_i$ and of the density perturbations
$\delta_i$ of the various components in a $\Lambda$CHDM model
having  $\Omega_b=0.023$, $\Omega_{\Lambda}=0.67$, $\Omega_m=0.33$,
$h=0.65$, $m_{\nu}=m_{\nu_{\mu}}+m_{\nu_{\tau}}=0.6$eV and
$\xi_{\nu}=\xi_{\nu_{\mu}}+\xi_{\nu_{\tau}}=5$.
 (a) and (a1): $\sin^2 2\theta_0=0.8$, $\Delta \xi_{\nu}=4$,
$\Delta m^2=0.24$eV$^2$ (continuous line) and
$\Delta m^2=0.18$eV$^2$ (dashed line).
(b) and (b1): $\Delta m^2=0.24$eV$^2$, $\Delta \xi_{\nu}=4$,
$\sin^2 2\theta_0=1$ (continuous line)
and $\sin^2 2\theta_0=0.45$ (dashed line).
Panels (c) and (c1): $\Delta m^2=0.24$eV$^2$, $\sin^2 2\theta_0=0.45$,
$\Delta \xi_{\nu}=4$ (continuous line) and
$\Delta \xi_{\nu}=2$ (dashed line).}
\vspace{-0.5cm}
\end{figure}
\begin{figure}
\vspace{-0.6cm}
\epsfig{figure=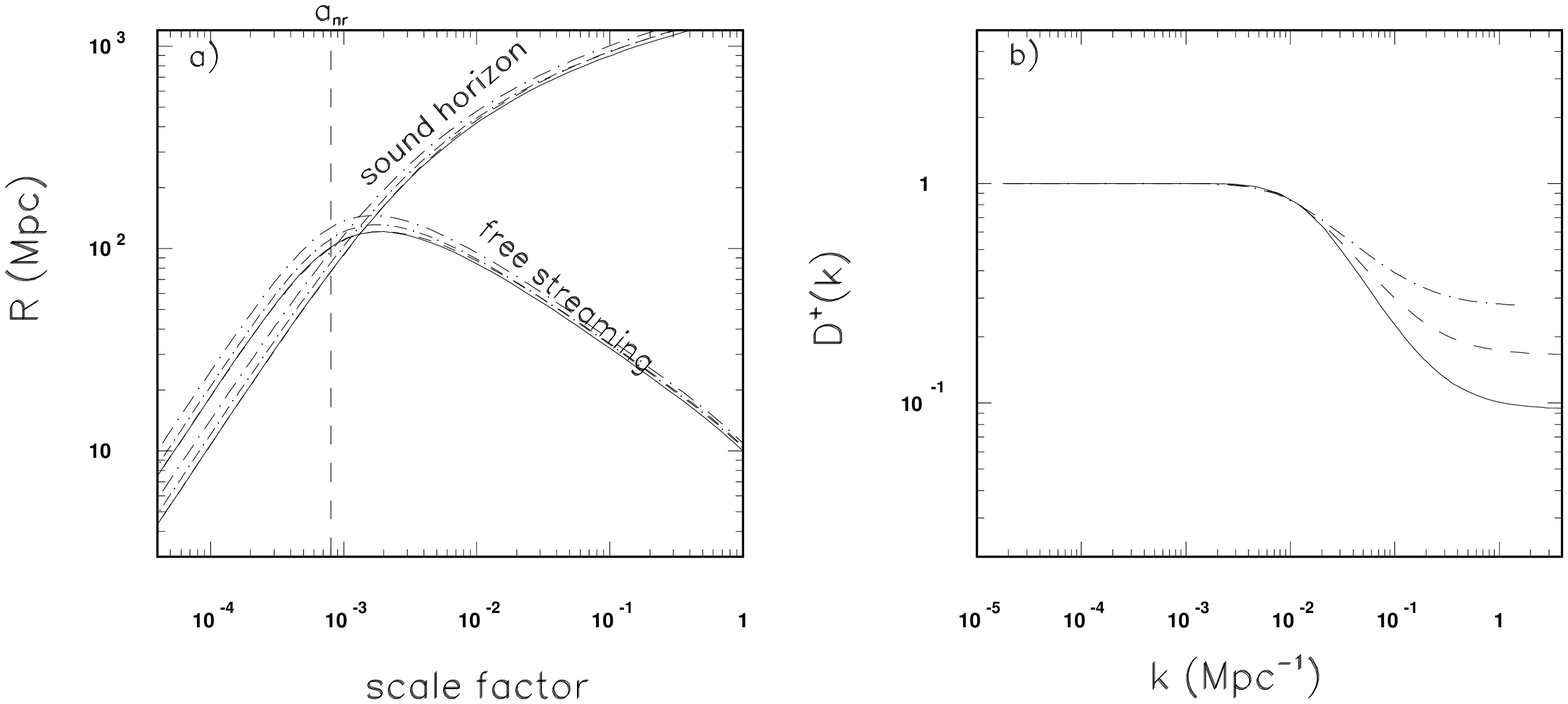,height=8cm,width=13cm}
\vspace{-0.3cm}
\caption{(a): The dependence of the time evolution of neutrino
free streaming distance and sound horizon distance on
$\Delta m^2$ and $\Delta\xi_{\nu}$ when:
$\Delta m^2=0.24$eV$^2$ and $\Delta\xi_{\nu}=4$ [$\Omega_{\nu}=0.16$]
(continuous line),
$\Delta m^2=0.2$eV$^2$ and $\Delta \xi_{\nu}=4$ [$\Omega_{\nu}=0.15$]
(dashed line),
$\Delta m^2=0.24$eV$^2$ and $\Delta \xi_{\nu}=1$ [$\Omega_{\nu}=0.07$]
(dash-dotted line).
The vertical line indicates the approximated
value of the scale factor when massive neutrinos start
to make the transition
from radiation to matter.
(b): The growth functions of the perturbations
at the present time ($z=0$) obtained for:
$\Delta m^2=0.24$eV$^2$ and $\Delta\xi_{\nu}=4$ [$\Omega_{\nu}=0.16$]
(continuous line),
$\Delta m^2=0.2$eV$^2$ and $\Delta \xi_{\nu}=4$ [$\Omega_{\nu}=0.15$]
(dashed line),
$\Delta m^2=0.24$eV$^2$ and $\Delta \xi_{\nu}=1$ [$\Omega_{\nu}=0.07$]
(dash-dotted line).
The cosmological model is a flat
$\Lambda$CHDM model
having: $\Omega_b=0.023$, $\Omega_{\Lambda}=0.67$, $\Omega_m=0.33$,
$h=0.65$, $m_{\nu}=m_{\nu_{\mu}}+m_{\nu_{\tau}}=0.6$eV and
$\xi_{\nu}=\xi_{\nu_{\mu}}+\xi_{\nu_{\tau}}=5$. }
\end{figure}
\begin{figure}
\vspace{-0.3cm}
\epsfig{figure=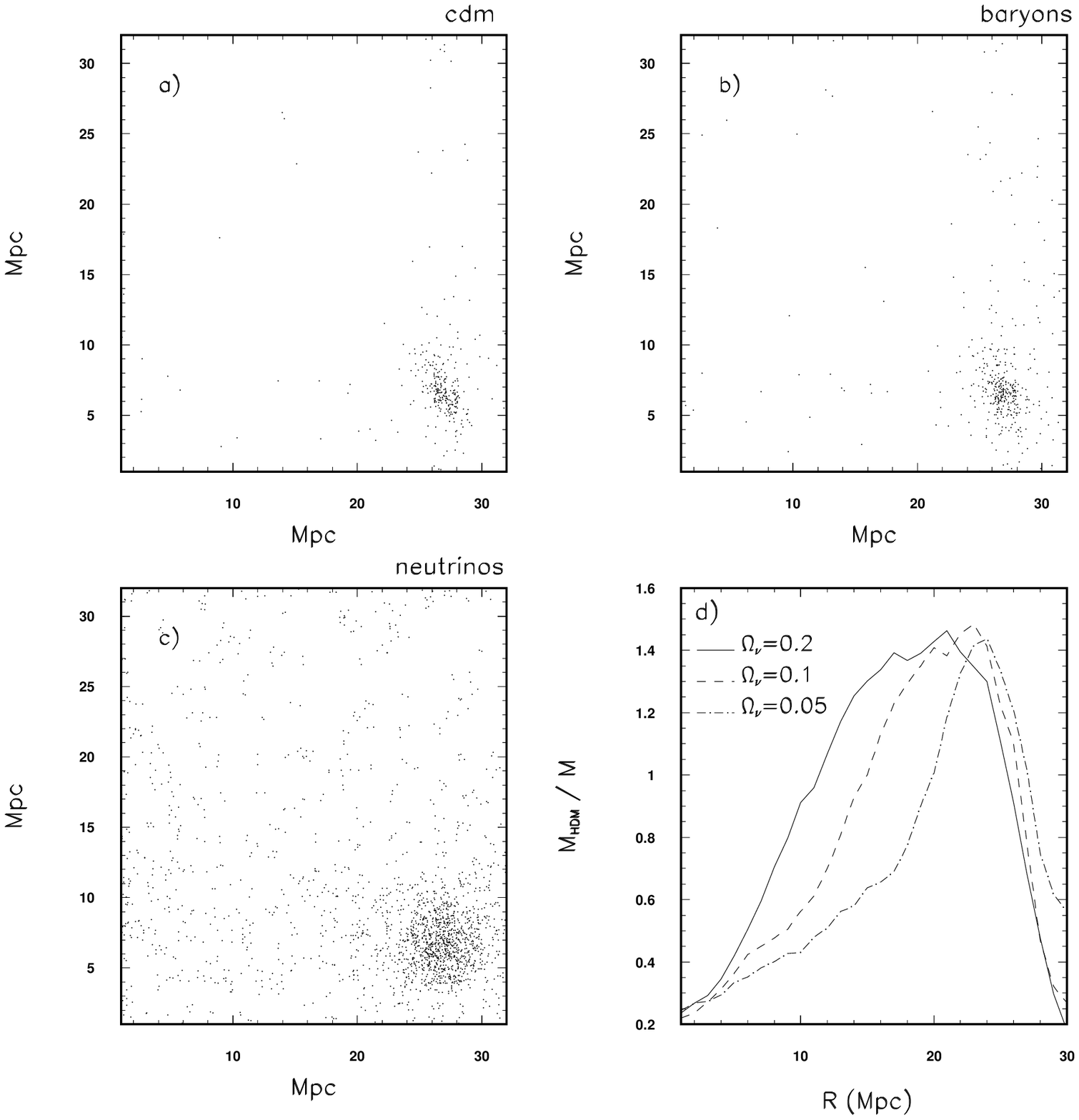,height=12cm}
\caption{ Maxima of the mass distribution of the
cold dark matter component (panel a),
baryonic component (panel b)
and hot dark matter component (panel c)
obtained from numerical simulations
of $32^3$ particles in a box of size $32$ Mpc
for the $\Lambda$CHDM model [$\Omega_b=0.023$,
$\Omega_{\Lambda}=0.67$, $\Omega_m=0.33$,
$h=0.65$, $m_{\nu}=m_{\nu_{\mu}}+m_{\nu_{\tau}}=0.6$eV and
$\xi_{\nu}=\xi_{\nu_{\mu}}+\xi_{\nu_{\tau}}=5$].
Panel d): The mass distribution of the HDM
component  obtained from numerical simulations for different
neutrino fractions (see also the text).
The total mass in all simulations is
$M=2.27\times10^{15}M_{\odot}$.}
\end{figure}

\section{CMB temperature and  matter density fluctuations}

During the epochs where the anisotropies are generated,
neutrinos with  masses in eV range
can have significant interactions
with the photons, baryons and cold dark matter particles only
via gravity. Thus differences introduced by the neutrino mixing
of the phase space distributions and the
lepton asymmetry of the
neutrino background are due to the differences in the gravitational
field of the neutrinos. As it was shown in the previous section,
the neutrino energy density and pressure as well as
the perturbed the energy density and pressure,
the shear stress and energy flux are modified
by the mixing of the neutrino degenerated background.
They determine differences in the equation
of state of massive neutrinos ($w=p_{\nu}/\rho_{\nu}$),
in the expansion law of the universe (eq. 4) and
originate the metric perturbations
affecting the growth of fluctuations in other components.

\subsection{The neutrino gravitational infall}

Characteristic to  the cosmological models involving
massive neutrinos is the scale-dependence
of the growth rate of perturbations.
Neutrinos can cluster via gravitational instabilities
only on distances below a characteristic distance, the
free streaming distance, analogous to the Jeans scale
of self-graviting systems (see e.g. \cite{MaB95}
and the references therein).\\
The neutrino free streaming distance is related to the causal
horizon distance $\eta(a)$ through \cite{dod95}:
\begin{eqnarray}
R_{fs}=\frac{1}{k_{fs}}=\frac{\eta(a)}{\sqrt{1+(a/a_{nr})^2}},
\hspace{0.3cm}
\eta(a)=\int^a_0 \frac{da}{a^2 H},
\end{eqnarray}
where: $k_{fs}$ is the free streaming wave number,
$a$ is the cosmological scale factor, $H$ is the
Hubble expansion rate given by the equation (4) and $a_{nr}$ is
the scale factor
when massive neutrinos start to become non-relativistic
($a_{nr}=(1+z_{nr})^{-1}\approx 3k_{B}T_{\nu_{0}}/ m_{\nu}c^2$).\\
Another important scale is given by the sound horizon distance of the
baryon-photon fluid defined as \cite{dod95}:
\begin{eqnarray}
R_s= \int^{\eta(a)}_0 c_s(a)d\eta, \hspace{0.3cm}
c_s^2(a)=\frac{1}{3}\frac{1}
{1+\frac{3}{4}a\frac{\Omega_{\gamma}}{\Omega_b}},
\end{eqnarray}
where: $c_s(a)$ is the sound speed and $\Omega_{\gamma}$ and $\Omega_b$
are the energy density parameters of photons and baryons at the present time.
Panel a) in  Figure 3
presents the dependence of time evolution of the neutrino
free streaming distance and sound horizon distance on
$\Delta m^2$ and $\Delta \xi_{\nu}$.
Panel b) of the same figure shows the evolution with the wave number
of the growth functions of perturbations obtained
for the same values of these parameters.
The increase of $\Delta m^2$ when $\Delta \xi_{\nu}$ is kept constant
leads to the increase of $\Omega_{\nu}$ and of $z_{nr}$.
The same effect is obtained when $\Delta m^2$ is kept constant and
$\Delta \xi_{\nu}$ is increased.
At early times, when neutrinos are relativistic,
the free streaming distance is approximately
the sound horizon distance. After the neutrinos become non-relativistic,
the free streaming distance decreases with time becoming
smaller than the sound horizon distance. Also,
the growth of perturbations is suppressed at smaller scales,
the magnitude of the suppression depending on neutrino parameters.\\
It follows that neutrinos can cluster gravitationally
on increasingly small scales at latter times, damping the amplitude
of the density perturbations.
This implies
that the characteristics of the cosmological structures at
small redshift,
as the mass density profile and the velocity dispersion profile of clusters,
can probe the neutrino properties.
Figure 4 presents maxima of the mass distribution of
the cold dark matter component, the baryonic component,
and the hot dark matter component,
obtained from numerical simulations in a flat $\Lambda$CHDM model.
The cold dark matter and
the baryonic matter components accrete the hod dark matter
component via gravity. The gravitational potential governing
this process is given by the Poisson equation
$\nabla^2 \Phi=4 \pi G a^2{\bar \rho} \epsilon_m(k,a)$,
where ${\bar \rho}=\Omega_m \rho_{cr}$ is the averaged density,
$\rho_{cr}$ is the critical density and $\epsilon_m(k,a)$
is the net perturbation of the density field.
Also, panel d) of the Figure 4
presents the mass distribution of the HDM component
(obtained for each simulation bin
and normalized to the total mass of each bin) for different
neutrino fractions.
\begin{figure}
\vspace{-2cm}
\epsfig{figure=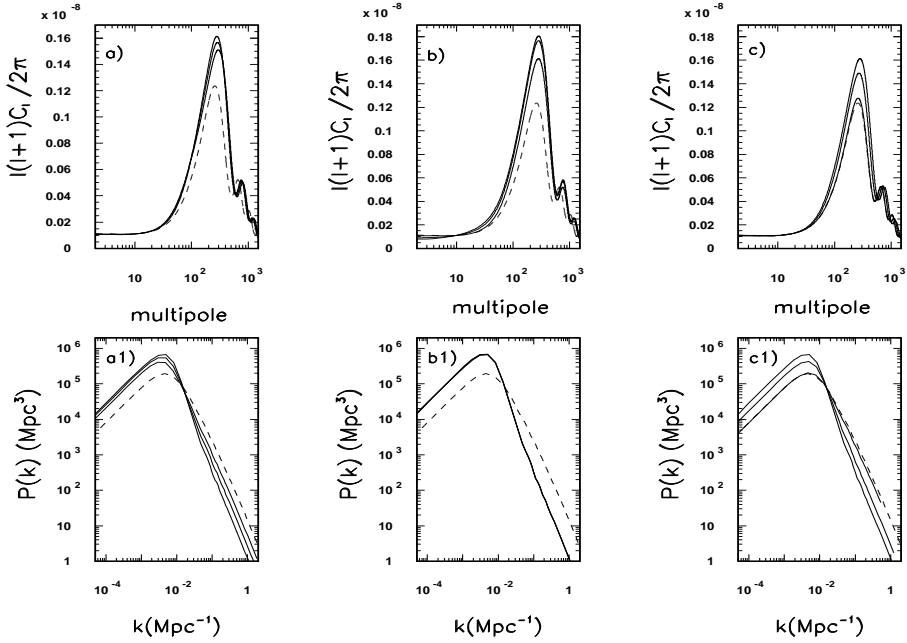,height=11cm,width=13cm}
\caption{The CMB anisotropy power spectra
and the matter density fluctuations power spectra.
Panels a) and a1):  $\sin^2 2\theta_0=0.8$, $\Delta \xi_{\nu}=4$ and
(from top to bottom) $\Delta m^2$=0.24, 0.2, 0.01 eV$^{2}$.
Panels b) and b1):$\Delta m^2=0.24$eV$^{2}$,  $\Delta \xi_{\nu}=4$
and (from top to bottom) $\sin^2 2\theta_0$=1, 0.8, 0.45.
Panels c) and c1):$\Delta m^2=0.24$eV$^{2}$, $\sin^2 2\theta_0$=0.8
and (from top to bottom) $\Delta \xi_{\nu}$=4, 3, 1.
In each panel are presented (dashed lines)
the power spectra obtained  in the same cosmological model
when $\Delta m^2=0$, $\Delta \xi_{\nu}=0$ and $\sin^2 2\theta_0=0$.}
\end{figure}

\subsection{CMB anisotropy and the matter power spectra}

In the linear perturbation theory, the CMB anisotropy
and  matter transfer function
are computed by the integration of coupled and linearized Einstein,
Boltzmann and fluid equations \cite{MaB95} that describe
the time evolution of the metric perturbations in the perturbed density
field and the time evolution of the density fields
for all the relevant particles.
Figure 5 presents the dependence of the CMB anisotropy
power spectra (left panels)
and matter density fluctuations
power spectra (right panels)
on the neutrino difference of the
squared masses  $\Delta m^2$, the mixing angle
$\sin^2 2 \theta_0$ and  the lepton asymmetry $L_{\nu}$ \cite{Popa2000}.
For all cases it is assumed a total neutrino mass
$m_{\nu}=m_{\nu_{\mu}}+m_{\nu_{\tau}}=0.6$eV and a total neutrino
degeneracy parameter $\xi_{\nu}=\xi_{\nu_{\mu}}+\xi_{\nu_{\tau}}=5$.
The CMB power spectra are normalized to COBE/DMR four-Year
data and the matter density fluctuations power spectra are
normalized to the abundance of the rich clusters at
the present time \cite{Popa2001}.\\
The CMB anisotropy power spectra show two distinct features:
a vertical shift of the $C_l$ at large $l$ (present in all panels)
with the increasing of $\Delta m^2$, $\sin^2 2\theta_0$ and
$\Delta \xi_{\nu}$ values,
that results from the modification of neutrino free streaming scales, and
an horizontal shift of the Doppler peaks to lower $l$ when
$\Delta m^2$ and $\Delta \xi_{\nu}$
are increased,
that results from the increase of the sound horizon at recombination.
The horizontal shift of $C_l$ is not present in panel b1)
because the variation of $\sin^2 2 \theta_0$, when $\Delta m^2$ and
$\Delta \xi_{\nu}$ are fixed,
does not change $\Omega_\nu$ and consequently the Hubble
expansion rate is not changed.
The matter density fluctuations power spectra reflects the
same features: the increasing of $\Delta m^2$ and
$\Delta \xi_{\nu}$ values
results in the modification of neutrino free streaming scales
and consequently a suppression of the growth of
fluctuations on all scales below the free streaming scale
($k_{fs}\simeq 0.02$Mpc$^{-1}$).
By using the Fisher information matrix method
it was found \cite{Popa2000} that
the future high precision CMB experiments as {\sc Planck}
surveyor
will permit to detect a lepton asymmetry
of the neutrino background
$\L_{\nu} \geq 4.5 \times 10^{-3}$,
a difference of the neutrino squared masses
$\Delta m^2 \geq 9.7 \times 10^{-3}$eV$^2$ and a mixing angle
$\sin^2 2\theta_0 \geq 0.13$.
Also, the combined data from {\sc Planck} and Sloan Digital Sky Survey
will permit the determination of the total neutrino mass
$m_{\nu} \geq 3\times 10^{-2}$eV,
with any prior information on cosmological parameters
\cite{Popa2001}.

\section{Conclusions}

The standard Cold Dark Matter model has difficulties
in matching the CMB and LSS measurements.
Cosmological
models involving a mixture of CDM and HDM particles, the CHDM models,
are able to fit,
the  excess large scale power seen in galaxy surveys
and the CMB temperature fluctuations.

The results from neutrino oscillation experiments indicate a non-zero
neutrino mass in the eV range, or lower. This  implies a non-negligible
contribution of neutrinos to the total energy density
of the universe.
The experimental evidence indicating a present
low matter density universe,  dominated by the vacuum energy
($\Omega_m \simeq 0.3$, $\Omega_{\Lambda}\simeq 0.7$) and a higher
Hubble parameter value ($H_0 \simeq 65$ km s$^{-1}$Mpc$^{-1}$)
are in agreement with cosmological models involving neutrinos
if one considers the lepton asymmetry of the neutrino background.

At the times when the anisotropies were generated, neutrinos
had significant interactions with the photons, baryons and cold
dark matter particles only via gravity.
Neutrinos can not cluster via gravitational effects on scales $k$
below the free streaming scale $k<k_{fs}$ (large wavelengths).
For $k>k_{fs}$ the growth of the density perturbations is suppressed,
the magnitude of this suppression depending on $\Omega_m$,
$\Omega_{\nu}$, $\Delta m^2$ and $\Delta \xi_{\nu}$.

The neutrino homogeneous quantities (density and
pressure) and inhomogeneous quantities
(density and pressure perturbations,
the shear stress and the energy flux)
are also changed in the presence of non-degenerated neutrino mixing,
leaving imprints on the CMB temperature fluctuations and
the matter density fluctuations power spectra.

The determination of the fundamental cosmological parameters
with high precision CMB and LSS surveys requires complementary knowledge
of neutrino properties from cosmic rays and long base-line experiments.

\vspace{0.6cm}
{\bf Acknowledgements:} I acknowledge  the organizers
and  V.~Berezinsky for useful discussions during this
workshop.

\end{document}